**Title:** Monitoring Deployed AI Systems in Health Care

**Authors**: Timothy Keyes, PhD*; Alison Callahan, PhD, MIS*; Abby S. Pandya, MBA, MS*; Nerissa Ambers, MPH; Juan M. Banda, PhD; Miguel Fuentes, MS; Carlene Lugtu, MCiM, BSN, RN; Pranav Masariya, MS; Srikar Nallan, MS; Connor O'Brien, MS; Thomas Wang, PhD; Emily Alsentzer, PhD; Jonathan H. Chen, MD, PhD; Dev Dash, MD, MPH; Matthew A. Eisenberg, MD; Patricia Garcia, MD; Nikesh Kotecha, PhD; Anurang Revri, MS; Michael A. Pfeffer, MD; Nigam H. Shah MBBS, PhD**; Sneha S. Jain, MD, MBA**

* These authors contributed equally as co-first authors.

**These authors contributed equally as co-last authors.

**Funding**: J.C. has received research funding support in part by NIH/National Institute of Allergy and Infectious Diseases (1R01AI17812101); NIH-NCATS-Clinical & Translational Science Award (UM1TR004921); Stanford Bio-X Interdisciplinary Initiatives Seed Grants Program (IIP); NIH/Center for Undiagnosed Diseases at Stanford (U01 NS134358); Stanford RAISE Health Seed Grant 2024; Josiah Macy Jr. Foundation (AI in Medical Education)

**Acknowledgements:** Danton Char, Clancy Dennis, Duncan McElfresh, Vishantan Kumar, Michelle Mello, Shyon Parsa, Eduardo Perez Guerrero, Patrick M. Sculley, Aditya Sharma, Margaret Smith.

AI tools were used to aid in the editing of this manuscript. These tools were used under human oversight, and all scientific content, analysis, and conclusions reflect the authors' original work and judgment. All final content was determined and approved by the authors.

**Disclosures**: A.C. is an advisor to Atropos Health. J.H.C. is a co-founder of Reaction Explorer LLC that develops and licenses organic chemistry education software and discloses paid medical expert witness fees from Sutton Pierce, Younker Hyde MacFarlane, Sykes McAllister, Elite Experts; consulting fees from ISHI Health; paid honoraria or travel expenses for invited presentations by insitro, General Reinsurance Corporation, Cozeva, and other industry conferences, academic institutions, and health systems. N.H.S reports being a cofounder of Prealize Health (a predictive analytics company) and Atropos Health (an on-demand evidence generation company); receiving funding from the Chan Zuckerberg Institute for developing classifiers for rare diseases; and serving on the Board of the Coalition for Healthcare AI (CHAI), a consensus-building organization providing guidelines for the responsible use of artificial intelligence in health care. E.A. reports consulting fees from Fourier Health. N.H.S serves as a scientific advisor to Opala, Curai Health, JnJ Innovative Medicines, and AbbVie pharmaceuticals. S.S.J reports consulting fees from Bristol Myers Squibb, ARTIS Ventures, and Broadview Ventures outside of the submitted work. The remaining authors report no relevant disclosures or competing interests.

**Address for correspondence**:

Sneha S. Jain, MD, MBA

Center for Academic Medicine

Department of Medicine, Division of Cardiovascular Medicine

Stanford University School of Medicine

453 Quarry Road

Palo Alto, CA 94304

e-mail: snehashahjain@stanford.edu

## Abstract

Post-deployment monitoring of artificial intelligence (AI) systems in health care is essential to ensure their safety, quality, and sustained benefit—and to support governance decisions about which systems to update, modify, or decommission. Motivated by these needs, we developed a framework for monitoring deployed AI systems that is organized around three complementary principles: system integrity, performance, and impact. *System integrity monitoring* focuses on maximizing system uptime, detecting runtime errors, and identifying when changes to the surrounding IT ecosystem have unintended effects. *Performance monitoring* focuses on maintaining accurate and equitable system behavior in the face of changing health care practices (and thus input data) over time. *Impact monitoring* assesses whether a deployed system continues to have value in the form of benefit to clinicians, staff, and patients. Drawing on examples of deployed AI systems at our academic medical center, we provide practical guidance for creating monitoring plans based on these principles that specify which metrics to measure, when those metrics should be reviewed, who is responsible for acting when metrics change, and what concrete follow-up actions should be taken—for both traditional and generative AI. We also discuss challenges in implementing this framework, including the effort and cost of monitoring for health systems with limited resources as well as the difficulty of incorporating data-driven monitoring practices into complex organizations where conflicting priorities and definitions of success often coexist. This framework offers a practical template and starting point for health systems seeking to ensure that AI deployments remain safe and effective over time.

## Motivation and Background

Effectively using AI in health care demands more than performant AI systems; it requires a governance process to decide which AI systems to deploy and when to refine, replace, or retire them. Post-deployment monitoring that is *actionable* is necessary for such governance, providing clear specification of what should be measured, at what cadence, who is responsible for responding, and how they should do so. Without governance—and monitoring to support it—errors such as AI tools inviting patients to the wrong screening[1] and poor model performance going unaddressed[2] are bound to occur.

Our goal is to provide a practical guide for monitoring deployed AI systems based on their design and behavior, the workflow(s) into which they are integrated, and their intended effects. Prior work[3–7] has described statistical tests, deployment and integration patterns, and other technical processes for monitoring AI systems (e.g. control charts, model registries, continuous integration/continuous delivery pipelines, and dashboarding tools). Health care organizations and consortia have also recently proposed governance frameworks that provide high-level guidance on topics related to post-deployment monitoring of AI systems (such as AI system risk categorization, evaluation, and maintenance), but these efforts offer limited guidance for how monitoring should be operationalized in practice[8–10]. Our work addresses this gap by providing a pragmatic monitoring framework—with multiple concrete examples—focused specifically on enabling actionable decision-making by health system leadership. We ground this framework in our institutional experience developing monitoring plans for deployed AI systems.

At Stanford Health Care (SHC), post-deployment monitoring is embedded within a broader governance process called the "Responsible AI Lifecycle" (RAIL) to oversee the approval, resource allocation, and deployment of AI systems[11]. Established in 2023, RAIL codifies institutional workflows for requirements gathering; risk tiering; and Fair, Useful, Reliable Model (FURM) assessments to ensure all deployed systems meet rigorous ethical and technical standards. Creation of a system-specific monitoring plan is a core output of FURM assessments, with ethics findings—including from interviews with a patient panel that surfaces their perspectives on the AI system's use—informing subgroup-specific monitoring when appropriate. Since 2022, TDS has conducted FURM assessments of 21 AI systems, both sold by vendors and developed in-house. Of these, 3 systems were assessed and deployed prior to the development of our monitoring framework, 5 systems were assessed and not deployed based on our assessment, and the remaining 13 have system-specific monitoring plans to enable regular review and inform decisions to modify or decommission tools that may no longer be useful. How we developed these monitoring plans is the focus of this article – specifically, the approach we use for defining what to monitor, the methods and tools we have developed to do the monitoring, and how we partner with clinical and operational teams to identify who should take what action, and when, based on the readouts from monitoring.

## Why monitor?

Deploying AI systems in health care is an ongoing operational commitment. While selecting which models to deploy (and how) are critical steps for AI adoption within a health system, sustained benefit requires continual measurement of how well an AI system works and the continued verification of its usefulness[12]. Post-deployment AI monitoring must be action-oriented: when an AI system stops working as expected, we may need to act by fixing a broken data pipeline, retraining a predictive model, re-prompting or re-configuring a large-language model (LLM), or retiring a tool when it is no longer valuable.

This stance is motivated by the fact that deployed AI systems sit within a complex ecosystem of clinical applications, data pipelines, and third-party integrations. For example, SHC runs over 1500 software applications with nearly 3100 interfaces. Electronic health record (EHR) platforms undergo regular upgrades and perpetual optimization, and integrated systems can be updated or replaced with far-reaching effects on their downstream dependencies[13]. These types of changes can result in an AI system's failure to locate its input data (e.g. a feature table moves or a note type is renamed) or in its failure to deliver an output where it is expected (e.g. an API endpoint changes and predictions no longer post to their intended destination)[14,15]. Thus, AI system monitoring must continuously verify the end-to-end functionality of the system and its associated data pipelines so that these kinds of failures can be quickly remediated.

A second reason for monitoring a deployed AI system is that the statistical relationships that a system relies on rarely remain stable over time. For *traditional AI systems*—AI systems that have been trained to perform a specific task like predicting the onset of a disease or classifying patients into distinct risk categories—differences between development and deployment populations, evolving clinical practice, and changing documentation habits can change the relationships between a model's inputs and outputs[16]. This phenomenon (often called "dataset shift" or "concept drift") is well-described in the clinical informatics literature and often results in a gradual erosion of an AI system's accuracy over time[17,18]. While *generative AI systems*—AI systems like LLMs that have been trained on a large corpus of data to perform a diverse set of tasks, such as summarization or information extraction—may be more robust to this phenomenon than traditional AI systems, they often suffer from the same limitations[19]. They also present unique challenges. For example, due to the inherent flexibility of both the inputs and outputs of LLMs, use cases and prompting patterns can also evolve over time as users develop new prompts for novel tasks. These changes may expose additional failure modes that were neither evaluated nor anticipated at the time of deployment.

Third, AI systems are only useful if required personnel, equipment, and work capacity to execute a downstream workflow exist[20]. Therefore, monitoring must maintain a line of sight from system outputs to downstream actions and their outcomes over time to guide the decision to redesign a workflow, retrain users, or retire a tool.

Together, these considerations motivate the organization of our monitoring framework around three complementary principles—*system integrity*, *performance*, and *impact*—intended to ensure that AI systems remain technically sound, produce high-quality outputs, and deliver intended benefits in practice, respectively (**Figure 1**). The first and second of these principles derive from the field of machine learning operations (MLOps), the discipline of building, deploying, and governing machine learning systems in production[21]. The third is rooted in the principles of quality improvement (QI) and business intelligence (BI)[22].

*System integrity monitoring* indicates whether the AI system is running as expected and encompasses infrastructure and data pipeline functionality. *Performance monitoring* indicates whether the model underlying the AI system is accurate and equitable in its output over time (i.e. is not negatively impacted by changes to the practice of medicine, documentation patterns, and patient population, as described above). *Impact monitoring* indicates how the AI system is affecting downstream processes and their outcomes; depending on the workflow(s) into which the AI system is integrated, these may be health care processes and outcomes (e.g. treatments provided by a doctor and their effect on patients) or operational processes and outcomes (e.g. documentation and the time required to complete it).

## How to monitor

### *Overview*

Monitoring strategies differ based on whether an AI system is traditional or generative, and—among generative AI systems—whether interaction with the tool occurs via "fixed" or "open" prompts. In *fixed-prompt* systems, a single, standardized prompt is executed on eligible patients as a scheduled batch or in response to a specific trigger. Thus, end-users only see the LLM-generated output and cannot directly prompt the system themselves. An example of a fixed-prompt system is SHC's Inpatient Hospice LLM Screen (**Figure 2**, Row 7), which screens critically ill patients for a palliative medicine consult using eligibility criteria described in a fixed prompt. By contrast, *open-prompt* systems—like SHC's EHR-integrated chatbot user interface (UI) ChatEHR (**Figure 2**, Row 10)[23]—give clinicians interactive access to the LLM, allowing them to compose their own prompts and receive diverse responses. Accordingly, this framework tailors monitoring plans to the type of AI system (traditional vs. generative) and, for generative systems, to the interaction mode (fixed- vs. open-prompt), with distinct objectives and metrics for each.

Organizing AI monitoring around system integrity, performance, and impact has enabled SHC to implement comprehensive monitoring plans for 13 active deployments (6 traditional AI, 7 generative AI) and to design monitoring plans for 4 planned deployments (all generative AI). We describe our portfolio of deployed AI systems in **Figure 2** and provide abbreviated monitoring plans for a representative subset of them in **Table 1**.

*Tools and platforms for monitoring*

Whenever possible, we leverage data platforms that our IT group already uses to implement monitoring plans, rather than adding point solutions for specific deployments. This reduces integration debt and ensures that monitoring reports, dashboards, and alerts can be managed easily by the teams who maintain the AI system. For example, for Epic Cognitive Computing models, we use Epic's Model and Feature Management activity and Radar dashboards[24] to track monitoring metrics over time, enabling in-workflow monitoring by the Epic configuration teams who manage these deployments. For AI systems developed in-house, we use Databricks[25] dashboards to visualize the health of both traditional and generative model-serving REST APIs, statistical performance metrics over time, and deployment-specific downstream Key Performance Indicators (KPIs). Across all of SHC's AI deployments, ServiceNow[26] serves as the common intake location for user-reported incidents and change requests.

*System integrity monitoring*

System integrity monitoring detects whether AI model-serving pipelines run end-to-end with high availability, on the expected data, and with acceptable latency. For traditional AI systems, system integrity monitoring emphasizes local infrastructure and data pipeline functionality because models are typically deployed on health system IT resources; key endpoints include the frequency with which required inputs are available when the system is called, outputs are produced, and warnings or errors occur. For generative AI systems that often rely on externally maintained LLM APIs, the same endpoints apply with added attention to system availability and responsiveness. Across both traditional and generative AI deployments, we track the following metrics: service uptime/outages, mean API request latency (for latency-sensitive deployments), and failures in data retrieval (such as feature missingness for traditional AI models and text retrieval failures for LLM systems) or inference serving (such as API errors and timeouts).

Example system integrity monitoring plans are summarized in **Table 1**. In practice, these plans pair two complementary cadences of oversight: (1) *real-time support* that mobilizes engineers to remediate acute failures and (2) periodic *governance review*, in which longitudinal trends are assessed to flag deployments with persistently high failure rates that warrant corrective action or retirement (**Figure 3**, red boxes). Accordingly, prior to deployment we pre-specify metric-specific alert thresholds that, if exceeded, route notifications to the accountable technical team (data science, integrations, or application owners) for investigation and remediation. Furthermore, for generative AI deployments, calls to external LLMs are proxied through an LLM gateway (LiteLLM), which additionally captures request-level telemetry (latency, token counts, request/response size, and error codes) that can be aggregated to monitor API health across deployments.[27]

*Performance monitoring*

Performance monitoring assesses whether system outputs remain accurate with respect to specific statistical metrics over time and—when indicated by the ethics component of the corresponding FURM assessment—whether performance is consistent across patient subgroups. In SHC's current portfolio, deployed traditional AI systems consist only of two-class or multi-class predictors. Thus, the longitudinal metrics we compute include standard classification metrics, such as positive predictive value (PPV; also called precision), recall (sensitivity), specificity, and the area under the receiver operating characteristic curve (AUROC). In contrast, performance monitoring of generative AI systems focuses on the quality and relevance of model outputs and relies more heavily on user feedback than that of traditional AI systems. Regardless of the underlying AI type, monitoring AI system performance typically requires a strategy for obtaining "ground truth" labels against which to compare model output. For generative AI systems, this is often achieved via gold-standard, human-labeled benchmark datasets. Although LLM-as-a-judge and other silver-standard approaches are increasingly used when gold-standard labels are unavailable[28], their clinical validity remains uncertain and warrants cautious interpretation; accordingly, our current practice favors human-labeled reference sets whenever they are available.

Example performance monitoring plans for traditional and generative AI systems are provided in **Table 1**. Typically, these plans define each of the following:

1. The statistical metrics used to evaluate the AI system's accuracy or response quality
2. An approach for obtaining ground-truth labels to compute those metrics
3. Accountable reviewers and monitoring cadence (often monthly or quarterly, with subgroup analyses performed when indicated by the ethics assessment)
4. Performance bands and escalation criteria that trigger alerts for governance review and corrective action

For (4), alert thresholds are typically selected during monitoring plan development in collaboration with clinical and operational stakeholders, balancing three considerations: the anticipated clinical or operational consequences of degraded performance, the expected month-to-month statistical variability in performance metrics, and the organizational burden of investigating small changes in model performance that are likely attributable to noise. To reduce false alarms, we apply a persistence screen (e.g. multiple out-of-band months in a year), wherein consecutive deviations prompt immediate review, whereas nonconsecutive deviations are escalated during routine governance review. Performance monitoring alerts are intentionally designed on a longer time scale than system integrity alerts, as system integrity metrics generally have low background variance and clear, time-sensitive remediation requirements (e.g. restart a job, redeploy a container, etc.). Performance metrics, by contrast, are expected to fluctuate over time due to sampling variation and seasonal case-mix changes; accordingly, we require evidence of persistent performance metric degradation before committing to high-resource actions such as retraining or retirement.

Importantly, monitoring generative AI system performance entails additional considerations based on the system's interaction mode and intended use. For *fixed-prompt* generative AI deployments, we often use LLMs as zero-shot classifiers prompted to evaluate a patient's suitability for specific clinical or administrative actions. We therefore monitor performance longitudinally by linking model outputs to downstream clinical adjudication (captured in-workflow as structured data in the EHR) and computing task-specific metrics over time. Adjudicated cases accumulate into living benchmarks for each fixed-prompt system, supporting routine re-evaluation as LLM versions update and deprecate.[29] At SHC, this evaluation feedback loop is conducted at scale using MedHELM, an internally-developed framework that ingests gold-standard benchmark datasets and supports scheduled batch evaluations across model versions[30]. For *open-prompt* deployments like the ChatEHR UI (**Figure 2**, Row 10), exhaustive benchmarking is infeasible. We therefore use usage-informed benchmarking—using log analysis to identify common tasks and potential failure modes—to inform real-time safety guardrails for known unacceptable outputs (e.g. fabricated facts). Monitoring for open-prompt system performance remains an active area of research and development, and our approach will evolve as the field matures.

*Impact monitoring*

Impact monitoring assesses whether AI-guided workflows deliver its intended benefit after deployment, as reflected in measurable changes in downstream process and outcome metrics. For traditional AI systems, impact monitoring evaluates whether actions taken in response to model output translate into improvements in patient outcomes, operational efficiencies (labor and time savings), or health system finances (e.g. revenue generation, cost avoidance). For fixed-prompt generative AI systems deployed to automate specific tasks, impact monitoring is similar to that of traditional AI systems, with additional emphasis on tracking run-time cost, given that per-request pricing for frontier LLM APIs typically exceeds on-premises model-serving costs. For open-prompt generative AI systems, impact is assessed primarily via adoption and usage, under the assumption that a system that is heavily used is valuable to its users.

Across both traditional and generative AI systems, impact monitoring should include surveillance for unintended effects, including patient safety events, when appropriate. For example, in impact monitoring for our low-value laboratory test predictor (**Table 1**), we pair utilization and efficiency metrics (alert responses and reduction in repeat testing) with metrics intended to detect potential harm, including STAT laboratory ordering, rapid response calls, and inpatient mortality on deployed service lines. Identifying how to monitor for such unintended effects is typically accomplished via focused interviews with users and health system staff prior to deployment during the ethics component of an AI system's FURM assessment.

As shown in **Table 1**, in our monitoring plans, we specify both (i) process measures aligned to the workflow's decision points (e.g. orders placed, alert responses, time saved) and (ii) outcome measures appropriate to the use case. Metric selection, subgroup stratification (when indicated), and review cadence are informed by input from the operational and clinical teams that use the AI system, as well as the ethics component of the system's FURM Assessment. As with system integrity and impact monitoring, the monitoring plan predefines thresholds and recommended actions (e.g. workflow adjustment, model retraining/reconfiguring/retirement) prior to deployment. When deployments span multiple sites, metrics and thresholds are tied to site-specific targets and stakeholders to enable locally actionable governance.

**Ownership and action**

In addition to metrics tailored to properties of AI systems and how they are used, effective monitoring requires establishing the habits for review and response. We accomplish this by embedding monitoring metrics into existing operational rhythms and tools, thus avoiding the creation of parallel or siloed processes. For example, TDS modified our existing project management application to enable assigning clear ownership and tracking of required monitoring tasks.

We tie each monitoring metric and corresponding alert, dashboard, or report to a specific responsible individual, aligned with existing scope of responsibility whenever possible (**Figure 1**; "Personas"). We integrate and reinforce the review by including the monitoring attributes such as the owner, assigned group, cadence of review, and link to dashboards/reports in our ServiceNow Configuration Management Database (CMDB), a centralized system of records for all applications in our IT system (which includes an AI model inventory).

Monitoring also supports the overarching AI system post-deployment decision-making at three inflection points:

1. **Transitioning from silent to active deployment** involves connecting the AI system output to a live application or interface where end users can view and act on the information. This transition is typically guided by system integrity and performance metrics collected during silent deployment, as well as readiness of end users to engage with the tool.
2. **Conducting a 90-day post-go-live review** involves examining system integrity metrics after the AI system is live (and performance and impact metrics if relevant in the time frame) to confirm that the AI system is functioning reliably in production, without errors or disruptions. The review may identify actions (described below) required to stabilize the system.
3. **Sustaining operational relevance** involves conducting a review of monitoring metrics to assess whether the AI system is delivering its intended value and is aligned with business priorities, which may shift over time. This cadenced review may identify actions to retrain, reconfigure, or retire an AI system.

There are a variety of possible actions that may be taken based on readouts of system integrity, performance and impact monitoring (**Figure 3**). For example, system integrity monitoring can surface runtime failures requiring on-call remediation; performance monitoring can indicate declining model accuracy that requires retraining; and impact monitoring can motivate workflow redesign or retirement when an AI system has low user adoption. In practice, we have repeatedly acted on monitoring readouts across our deployed AI systems to enable organizational decision-making. For example, our post-go-live review of performance and impact metrics for our LLM-powered inpatient hospice screen supported its expansion to all inpatient units at SHC after a pilot deployment on 2 units. Impact review of the PAD risk classification model initially identified process metrics below specified thresholds, prompting workflow modifications to improve rates of PAD workup and identification. Monitoring review of six Epic-developed traditional AI systems resulted in retirements—two models (Likelihood of Unplanned Readmission version 1 and Risk of Patient No Show) were replaced with better-performing models, and four models (Risk of Inpatient Falls, ICU Length of Stay, ICU In-hospital Mortality Risk, and Risk of ICU Readmission or Mortality) were decommissioned because they were not connected to workflows and therefore could not support impact monitoring.

**Discussion**

Our experience developing and integrating post-deployment monitoring into SHC's Responsible AI Lifecycle demonstrates that our framework of system integrity, performance, and impact monitoring can enable impactful and timely action when AI systems do not behave as expected over time. A key strength of our approach is its robustness to variations in an AI system's underlying technology. With the advent of agentic AI and other emerging AI capabilities—whose adoption in medicine is rapidly approaching[31]—such a technology-agnostic approach is particularly important to ensure that monitoring efforts can keep pace with technological advancements. With new capabilities, novel monitoring challenges are certain to arise, and how to perform each component of monitoring will also need to evolve. For example, for agentic systems, performance of individual agents does not always translate to the performance of the end-to-end agentic system—thus creating challenges for performance monitoring[32].

Another important design feature of our approach is that we intentionally did not define equity as a fourth, standalone monitoring domain. Instead—consistent with organizational evidence that equity initiatives are most durable when embedded within routine institutional processes[33,34]—we incorporate equity as a cross-cutting requirement within both performance and impact monitoring through the measurement of subgroup-stratified metrics (as specified during the ethics assessment component of an AI system's FURM assessment). This design allows for differential benefits or harms to specific patient populations to be detected, escalated for governance review, and addressed through RAIL actions (workflow changes, retraining, or retirement).

As we implemented our monitoring framework across SHC's portfolio of deployed AI systems, we encountered several notable challenges. Some were expected and reflect the realities of introducing a new approach across a complex organization. For example, we identified a number of long-running AI systems that were monitored idiosyncratically or not at all. Harmonizing these variations in practice into a common schema required implementation of new tools as well as culture change and upskilling across many teams. Establishing a shared taxonomy via our monitoring framework provided the common language and structure to align expectations and map responsibilities to appropriate teams.

Furthermore, the number of deployed technology systems is often considered a tacit success metric for an IT group; thus, incorporating a monitoring framework that recommends long-term evaluation and potentially decommissioning some of those systems can be counter-cultural. At SHC, explicit governance processes and leadership support for retiring low-value tools mitigated this barrier.

Monitoring third-party solutions using our monitoring framework represented another challenge. For AI systems developed, maintained, and served by third-party vendors, it can be difficult to build effective monitoring solutions due to lower visibility into how they work and a limited ability to customize the metrics they make available for audit. This remains an active challenge—many vendors do not yet provide the telemetry necessary to align with our monitoring framework. For this reason, our current decision-making around monitoring third-party tools is primarily based on system integrity and impact metrics. We are also updating our FURM Assessment intake process to require that vendors disclose, upfront, what support they provide for bespoke monitoring capabilities so that our governance groups can incorporate this information into purchasing decisions. We see contractually requiring vendors to provide a minimal set of monitoring capabilities—including per-inference logging and secure APIs for exporting timestamped system inputs, outputs, and user-feedback—in enterprise software agreements as a potential path to address this challenge.

Our monitoring approach is not without limitations. Principal among these is that of resource intensity—sustaining comprehensive monitoring efforts requires dedicated data engineers, data scientists, product managers, clinical informaticians, and operations/business partners. Given the relatively modest IT budget of most health systems, many organizations may be unable to resource such efforts[31]. As the number and diversity of AI tools deployed across SHC grows, our monitoring processes will have to adapt to handle increased volume. To support this, we are establishing a triage process that tailors the depth of assessment and post-deployment monitoring to deployment scale and risk—considering factors like how many staff or patients it will impact, use in clinical decision-making, and the potential harms that could result from errors. Furthermore, acting on monitoring readouts often requires organizational change management (e.g. training, workflow redesign, and operational coordination), which additionally draws from the same finite pools of resources and personnel. One practical remedy for this problem is encapsulating our framework into software libraries and applications that automate most tasks, require minimal custom code, and can be disseminated within SHC and to peer institutions[32,33].

Furthermore, two technical limitations to our current monitoring approach merit explicit note. First, our default alerting threshold for performance monitoring (a 75-125% acceptance band relative to validation metrics) is a simple heuristic adopted to provide an expedient rule when explicit risk tolerances are difficult to prespecify. In principle, performance thresholds can and should be anchored in domain knowledge (e.g. choosing a minimally acceptable sensitivity or precision for an alert to be clinically useful), but in practice we have found that such thresholds are difficult to define prospectively in operational settings. Accordingly, we apply a conservative band intended to detect large, persistent deviations in model performance that clearly warrant action. As monitoring data accrue over time, we plan to replace this heuristic by estimating empirical performance variability for each deployment and transitioning to more rigorous statistical process control (SPC) or related control-chart methodologies, acknowledging that these methods still require parameter choices regarding what constitutes an actionable deviation[3]. Second, for generative AI systems, we currently rely on human-labeled benchmark datasets assembled through manual chart abstraction, which is labor-intensive and difficult-to-scale. Emerging strategies for semi-automated evaluation corpus construction and cautious use of LLM-as-a-judge silver-standards are exciting directions for the field that may enable higher-throughput monitoring that can be supplemented by targeted human review[35,36].

Looking ahead, we expect that we—and other health systems—will adopt an explicitly risk-based monitoring framework rather than assuming that all three components of monitoring are fully necessary for every deployment. Based on our initial experience, reasonable governance criteria for determining how to “right-size” monitoring practices for a given AI system may include dimensions such as whether the AI system functions as clinical decision support, whether it is patient-facing, and the number of steps between the AI system’s recommendation and downstream clinical action—drawing from risk frameworks articulated in the health law and ethics literature[36,37]. Health care IT operates in a highly regulated environment, governed by the Health Insurance Portability and Accountability Act (HIPAA), the Health Information Technology for Economic and Clinical Health (HITECH) Act, Centers for Medicare and Medicaid Services (CMS) billing requirements, and, when relevant, Food and Drug Administration (FDA) oversight of software-as-a-medical-device[37–39]. With AI now embedded across many health care applications, regulatory groups such as the Joint Commission are also introducing new guidance to promote safety, fairness, and accountability [41]. Monitoring frameworks will need to adapt as these regulatory requirements mature, translating evolving expectations into operational checks that support internal quality review and external compliance.

Both traditional and generative AI systems require unique monitoring considerations for deployment in clinical systems. Through experience implementing monitoring plans with concrete follow-up actions for 13 deployments, we demonstrate the capability for data-driven decision-making around the adoption, retraining, and retirement of AI tools. We share these as a holistic framework to guide such deployments that emphasizes actionable monitoring of AI system integrity, performance, and impact into governance processes.

**Figure captions**

**Figure 1 – The three anchoring principles of post-deployment AI monitoring.** Post-deployment AI monitoring can be organized into three complementary principles that apply to both traditional and generative AI systems. System integrity monitoring (top; red) verifies that IT infrastructure, data pipelines, and integrations are functional (high availability, acceptable latency, minimal downtime). Performance monitoring (middle; blue) evaluates the longitudinal accuracy and quality of AI system outputs to detect drift. Impact monitoring (bottom; green) verifies if the AI system produces sustained benefits to patients, health system staff, or health system finances over time. Together, these domains trigger corrective actions—such as repairing broken data pipelines, retraining or re-prompting models, or retiring tools—when problems cannot be remediated.

This figure provides a role- and metric-oriented schematization of our monitoring framework across these anchoring principles. Column 1 (Principle) names each anchoring principle, and Column 2 (Definition) states its objective. Column 3 (Personas) identifies the primary roles accountable for building and interpreting the metrics associated with each anchoring principle. Columns 4 and 5 provide example metrics for both traditional AI systems (Column 4) and generative AI systems (Column 5). Metrics are illustrative and should be tailored to each specific use case and deployment.

**Figure 2. Details about 13 traditional and generative AI systems deployed at Stanford Health Care.** This table provides details about 6 traditional AI and 7 generative AI systems

deployed and monitored at Stanford Health Care (SHC). For example monitoring plans selected from these deployments, see **Table 1**.

**Figure 3 - Action-oriented process diagram for post-deployment AI monitoring.** Following deployment, three parallel monitoring workstreams map to the anchoring principles outlined in Figure 1: System integrity (red), performance (blue), and impact (green). Each workstream specifies what to measure, when to act, and how to act: (i) system integrity—error/warning/failure logging and API telemetry trigger real-time on-call support; (ii) performance—statistical metric dashboards with threshold alerts reviewed at intervals defined during the pre-deployment FURM assessment; and (iii) impact—user feedback, outcomes, and process key performance indicators (KPIs) are organized into dashboards or reports that are included in periodic governance review. Metrics from all three workstreams route to governance committees for decisions to retrain, reconfigure, or retire the AI system, after which approved changes are implemented and monitoring continues.

# Figure 1

Anchoring Principles for AI System Monitoring

| Principle | Definition | Personas | Example Monitoring Metrics | |
|---|---|---|---|---|
| | | | Traditional AI | Generative AI |
| **System Integrity**<br>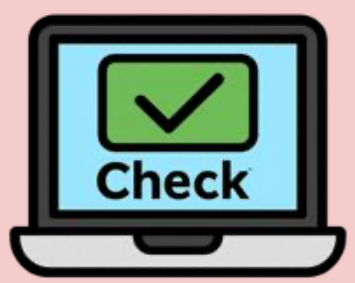<br> | Check that IT infrastructure, data pipelines, and integrations are functioning as intended to avoid extended AI system execution failures or downtime. | **Build:** Data/DevOps Engineer<br><br>**Interpret:** Analyst, Informaticist, or Data/DevOps Engineer | • Model uptime/downtime<br>• Missing data %<br>• Inference latency | • API uptime/downtime<br>• Token consumption rate<br>• Generation latency |
| **Performance**<br>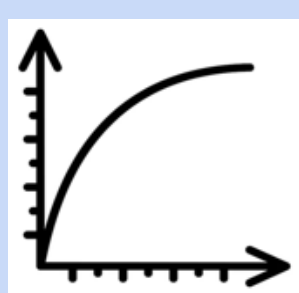 | Evaluate accuracy, quality, and safety of AI model outputs over time to detect drift or decay. | **Build:** Data Scientist/Engineer<br><br>**Interpret:** Analyst, Informaticist, or Data Scientist | • Statistical metrics (AUROC, precision, recall)<br>• Model Calibration<br>• Input/output drift detection | • Benchmark performance (e.g., MedHELM)<br>• Human-in-the-loop feedback<br>• Guardrail breaches |
| **Impact**<br>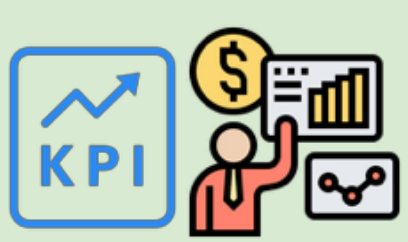<br> | Verify the continued benefits of the AI-guided workflow for patients, staff, and the organization. | **Build:** Business Intelligence Analyst, Data Scientist<br><br>**Interpret:** Analyst, Informaticist, or Clinical/Business Owner | • % Alerts acted on<br>• KPI improvements | • Usage/adoption<br>• KPI improvements (revenue generation, cost avoidance, labor savings)<br>• Time savings |

# Figure 2

| Name | Deployment Type | Developer | Brief Description |
|---|---|---|---|
| Peripheral Artery Disease (PAD) Classifier | Traditional | SHC | An XGBoost classifier of a patient's likelihood of having undiagnosed Peripheral Artery Disease (PAD). |
| Inpatient Advance Care Planning (ACP) | Traditional | SHC | A XGBoost predictor of an inpatient's likelihood of mortality in the next year used to guide early goals-of-care (advance care planning; ACP) conversations |
| Low-value Laboratory Test Predictor | Traditional | SHC | A neural network predictor of stable laboratory results to reduce low-value repeat testing by flagging routine lab orders that may be safely reduced or discontinued. |
| Intraoperative Transfusion Predictor | Traditional | SHC | A random forest predictor of the likelihood that a patient will need a blood transfusion on the day of surgery. |
| Likelihood Of Denial Recovery (Payment) | Traditional | Epic | A gradient-boosted tree predictor of the likelihood that an insurance denial will receive a future payment if appealed. |
| Likelihood of Unplanned Readmissions (V2) | Traditional | Epic | A random forest predictor of the likelihood that a patient will be readmitted to the hospital within 30 days of discharge from an inpatient admission. |
| Inpatient Hospice LLM Screen | Generative (Closed-prompt) | SHC | An LLM-powered workflow for detecting patients who may benefit from a palliative medicine consult for inpatient hospice care at end-of-life. |
| Surgical Co-management (SCM) Eligibility LLM Screen | Generative (Closed-prompt) | SHC | An LLM-powered workflow for identifying patients scheduled for surgery who might benefit from additional care from a hospitalist attending during post-surgical inpatient recovery. |
| AI-Augmented Notes Review for Surgical Site Infection (SSI) | Generative (Closed-prompt) | SHC | An LLM-powered workflow for identifying patients who may have experienced a surgical-site infection after surgery to assist in manual chart review for mandatory reporting. |
| ChatEHR Interactive User Interface (UI) | Generative (Open-prompt) | SHC | An EHR-integrated chatbot for searching patient charts using natural language. |
| (LLM-generated) Draft Denial Appeal Letters (Hospital Billing) | Generative (Closed-prompt) | Epic | An LLM-powered workflow for drafting an explanation for why a payer should reverse its decision to deny payment for services based on clinical documentation. |
| (LLM-generated) Draft Denial Appeal Letters (Professional Billing) | Generative (Closed-prompt) | Epic | An LLM-powered workflow for drafting an explanation for why a payer should reverse its decision to deny payment for services based on clinical documentation. |
| (LLM-generated) Medical Necessity Insights for Prior Authorizations | Generative (Closed-prompt) | Epic | An LLM-powered documentation assistant that reviews clinical notes, orders, and results to generate a payer-facing medical-necessity justification for treatment. |

# Figure 3

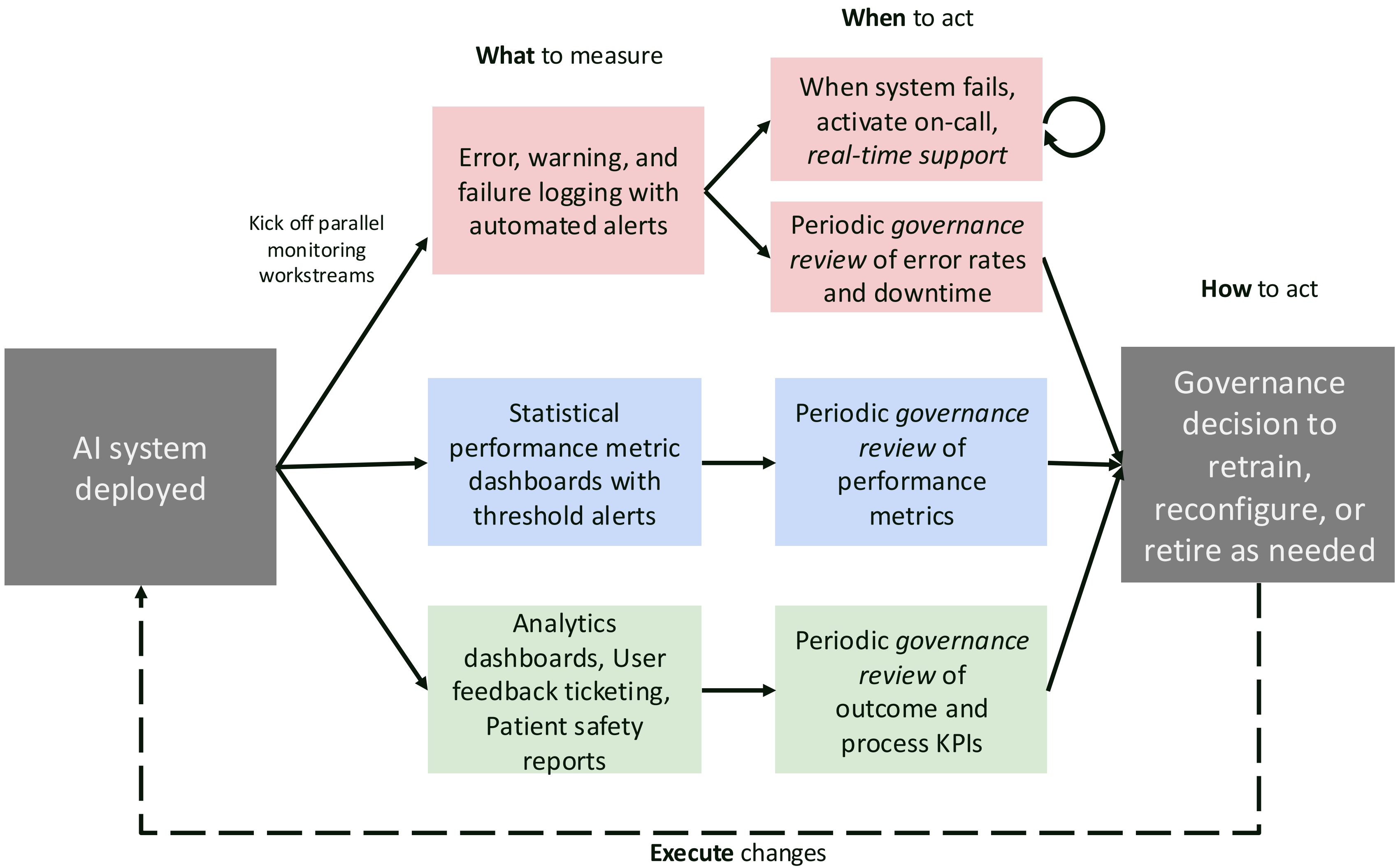

What to measure
When to act
How to act
AI system deployed
Kick off parallel monitoring workstreams
Error, warning, and failure logging with automated alerts
When system fails, activate on-call, *real-time support*
Periodic *governance review* of error rates and downtime
Statistical performance metric dashboards with threshold alerts
Periodic *governance review* of performance metrics
Analytics dashboards, User feedback ticketing, Patient safety reports
Periodic *governance review* of outcome and process KPIs
Governance decision to retrain, reconfigure, or retire as needed
Execute changes

Table 1 – Monitoring Plans for 10 deployed AI systems at Stanford Health Care.

| | System Integrity | | | Performance | | | Impact | | |
|---|---|---|---|---|---|---|---|---|---|
| **AI-Guided Workflow** | **Metrics** | **Action** | **Owner, Tool, Cadence** | **Metrics** | **Action** | **Owner, Tool, Cadence** | **Metrics** | **Action** | **Owner, Tool, Cadence** |
| **Peripheral Artery Disease (PAD) classifier**<br><br>System details: An XGBoost classifier of a patient's likelihood for having undiagnosed Peripheral Artery Disease (PAD).<br><br>Workflow details: The model is used to identify patients with a high risk of undiagnosed PAD in SHC's primary care population, flag these high-risk patients for PAD workup, and to increase the rate of appropriate treatment. | 1. Number of scores generated<br>2. Number of flagged patients (positive scores)<br>3. Number of inference errors/failures<br>4. Feature distribution<br>5. Prediction probability distribution | Real-time support: Send an alert when the inference job fails; respond to alerts and resolve pipeline or inference issues<br><br>Governance review: Compile a report summarizing the number and types of errors and incident reports for monthly review. | Real-time support: Data Science team resolves failures in real-time.<br><br>Governance review: Data Science team reviews Databricks Dashboard monthly. | The following metrics are computed on a rolling basis across the entire inference population over time:<br><br>1. Sensitivity<br>2. PPV<br><br>In addition, a report of these metrics across patient subgroups is generated annually. | If either metric deviates outside the performance band of 75-125% of pre-deployment values 3 or more months within a given year, resource data science effort to retrain.<br><br>If retraining does not improve performance, initiate model retirement. | Data Science team reviews Databricks Dashboard quarterly | Among all patients flagged by the model, we compute the following:<br><br>1. Number of patients who receive a PAD workup<br>2. Number of patients ultimately diagnosed with PAD<br>3. Detailed process metrics for each step of the PAD workup process (completion of pre-visit questionnaire, provider notification, referral scheduled, completed visits for diagnostic testing) | Manually review trends and assess continued value with the business owner and clinical team.<br><br>Adjust workflow or retire if business owner or clinical team feel that the tool is no longer useful | Business owner reviews report in Epic 3 months post-deployment, and annually thereafter |
| **Inpatient Advance Care Planning (ACP)**<br><br>System details: An XGBoost predictor of an inpatient's likelihood of mortality in the next year used to guide early goals-of-care (advance care planning; ACP) conversations<br><br>Workflow details: The model is used to identify patients who may benefit from an early goals-of-care conversation from the Serious Illness Care Program (SICP) | 1. Number of scores generated<br>2. Number of flagged patients (positive scores)<br>3. Number of inference errors/failures<br>4. Feature distribution<br>5. Prediction probability distribution | Real-time support: Respond to alerts and resolve pipeline or inference issues<br><br>Governance review: Compile a report summarizing the number and types of errors and incident reports for monthly review. | Real-time support: Data Science team resolves failures in real-time.<br><br>Governance review: Data Science team reviews Databricks Dashboard monthly. | The following metrics are computed on a rolling basis across the entire inference population over time:<br><br>1. Sensitivity<br>2. PPV<br><br>In addition, a report of these metrics across patient subgroups is generated annually. | If either metric deviates outside the performance band of 75-125% of pre-deployment values 3 or more months within a given year, resource data science effort to retrain.<br><br>If retraining does not improve performance, initiate model retirement. | Data Science team reviews Databricks Dashboard quarterly | The Serious Illness Care Program (SICP) tracks the following KPIs:<br><br>1. Number of patients flagged by the model.<br>2. Number of patients with documentation of a goals of care conversation.<br>3. Number of unique patients engaged by SICP.<br>4. The number of documented goals of care conversations per provider. | Review trends, assess continued value, adjust workflow or retire if needed. | Business owner review report monthly. |
| **Low-value Laboratory Test Predictor**<br><br>System details: A neural network predictor of stable laboratory results to reduce low-value repeat testing by flagging routine lab orders that may be safely discontinued.<br><br>Workflow details: The model is used to reduce the frequency of unnecessary and low value lab tests by alerting clinicians when routine lab orders are highly likely to be stable relative to previous values. | 1. Number/ proportion of patients with any missing data<br>2. Number/ proportion of patients with failed output generation<br>3. Prevalence of input feature elements<br>4. Prevalence of output events (flagged laboratory orders) | Real-time support: For metrics 1-2, an alert will fire if >10% of patients have any missing features or an inference error (missing features are expected to be very rare). Data Science must then identify the cause of missingness or failures and resolve data pipeline issues.<br><br>Governance review: For metrics 3-4, if there is a change > 1 standard deviation over historical values, an alert will be fired. Data Science must then investigate the source of drift and present to governance committee. | Real-time support: Data Science team responds to alerts from metrics 1-2 at a daily cadence.<br><br>Governance review: Governance review of alerts from metrics 3-4 quarterly; after 1 year of stable metrics, lengthen to an annual cadence. | The following metrics are computed on a rolling basis across the entire inference population over time:<br><br>1. AUROC<br>2. PPV<br>3. Sensitivity<br>4. Specificity<br>5. NPV | If any metric deviates outside a performance band of > 1 standard deviation of historical values, trigger an alert to present to project owner/clinical teams for decision to retrain, retire, or make other changes. | Quarterly, and at any version updates; after one year of stable metrics, lengthen to an annual cadence. | 1. Alert acceptance vs. override rates<br>2. Inpatient repeat lab order (within 24 hours) frequency<br>3. Patient lab collections per hospital day average<br>4. Adverse events: STAT lab orders, rapid response calls, ICU transfers, inpatient death per patient hospital day on deployed service lines<br>5. Eligible user feedback on system performance (qualitative) | If system is having desired impact, metrics 2-3 are expected to decrease without an increase in metric 4.<br><br>If this is not observed, if metric 1 drops >1 standard deviation over historical levels, or users give negative feedback (metric 5), present finding to governance team for decision to retrain, change workflow, or retire. | Business owner and clinical teams review metrics 1-4 quarterly and metric 5 annually. |
| **Likelihood of Unplanned Readmissions (version 2)**<br><br>System details: A random forest predictor of the likelihood that a patient will be readmitted to the hospital within 30 days of discharge from an inpatient admission. An | 1. Number of scores generated<br>2. Number of and proportion of flagged patients (positive scores)<br>3. Number of inference errors/failures | Real-time support: When any of metrics 1-4 exceed a 20% increase over the previous execution (Epic's recommended configuration), send alert to Applications team. | Real-time support: Application team resolves alerts in real-time.<br><br>Governance review: Application team reviews | 1. Sensitivity<br>2. Specificity<br>3. AUROC<br>4. AUPRC<br>5. Precision<br>6. C-statistic<br>7. Flag rate | Investigate deviations in any metric outside a performance band of 75-125% compared to that observed during model validation.<br><br>If this occurs, | Informatics reviews Model Inpatient Applications team reviews performance dashboard monthly and performs subgroup analysis yearly | 1. Readmission rate<br>2. Referral, Scheduled<br>3. Completed Visits | Review trends, assess continued value, adjust workflow or retire if needed | Business owner reviews Dashboards 6 months post-deployment and annually thereafter (for readmission rate) or monthly (for visit metrics) |
| Epic Cognitive Computing model.<br><br><u>Workflow details:</u> Used to schedule follow-up Primary Care appointments for high-risk patients before discharge. | 4. Feature distribution<br>5. Prediction probability distribution<br>1. | <u>Governance review:</u> Metrics 1-5 reviewed manually with business owner. | Epic Model Feature Management Dashboard monthly. | | retrain or retire as needed. | | | | |
| **Inpatient Hospice LLM Screen**<br><br><u>System details:</u> An LLM-powered workflow for detecting patients who may benefit from a palliative medicine consult from end-of-life inpatient hospice care.<br><br><u>Workflow details:</u> Used by inpatient Palliative Medicine APPs and nursing staff, who review a system-generated list of flagged patients and determine whether to reach out to the patient's care team to initiate a referral for hospice consultation. | 1. Number of eligible patients per daily execution<br>2. Number and proportion of flagged patients (positive scores)<br>3. Number of inference errors/failures | <u>Real-time support:</u> For metric 3, fire an alert if there are any inference errors for a single daily run. Respond to alerts and resolve pipeline or inference issues<br><br><u>Governance review:</u> For metric 1, send an alert if the number of eligible patients increases outside a 75-125% band around historical values. For metric 2, review with clinical stakeholders if the number of flagged patients regularly exceeds clinical capacity. | <u>Real-time support:</u> Integration team responds to alerts in real-time.<br><br><u>Governance review:</u> Data Science team reviews Databricks Dashboard monthly. | Calculate the number and proportion of flagged patients in each human feedback category:<br>• Reach out to team<br>• Continue to monitor<br>• Not relevant<br><br>Annually, generate a report breaking down the above into patient subgroups identified during ethics assessment. | Investigate when performance metrics deviate from baseline. If the proportion of flagged patients marked as "not relevant" grows beyond a tolerable threshold, consider reconfiguring the pipeline (e.g. changing the prompt) or retirement. | Data Science team reviews Databricks Dashboard monthly and Analytics performs subgroup analysis yearly. | 1. Total number of flagged patients (daily/weekly/monthly)<br>2. Distribution of flagged patients by feedback category<br>3. Of "outreach to team" patients: number of patients referred and admitted to IP Hospice<br>4. Generation cost over time<br>5. Number of potential misses, per monthly manual review | Monitor trends manually with clinical team and business owners.<br><br>If hospice enrollments fall below threshold or false negatives rise, reassess utility. | Business owner reviews report 3 months post-deployment, and annually thereafter.<br><br>Analytics creates specific ad hoc subgroup analysis yearly. |
| **Surgical Co-management (SCM) Eligibility LLM Screen**<br><br><u>System details:</u> An LLM-powered workflow for identifying patients sch eduled for surgery who might benefit from addition al care from a hospitalist attending during post-surgical inpatient recovery.<br><br><u>Workflow details:</u> Used by hospitalists to identity patients suitable for co-management during their hospital stay. | 1. Number of eligible patients per week/month<br>2. Number and proportion of flagged patients (positive scores)<br>3. Number of inference errors/failures | <u>Real-time support:</u> For metric 3, fire an alert if there are any inference errors for a single daily run. Respond to alerts and resolve pipeline or inference issues<br><br><u>Governance review:</u> For metric 1, send an alert if the number of eligible patients increases outside a 75-125% band around historical values. For metric 2, review with clinical stakeholders if the number of flagged patients regularly exceeds clinical capacity. | <u>Real-time support:</u> Integration team responds to alerts in real-time.<br><br><u>Governance review:</u> Data Science team reviews Databricks Dashboard monthly. | Calculate the number and proportion of flagged patients in each clinician feedback category:<br>• "Yes" (Plan to co-manage this patient)<br>• "No" (Do not plan to co-manage this patient)<br><br>Annually, generate a report breaking down the above into patient subgroups identified during ethics assessment. | Investigate when performance metrics deviate from baseline. If the proportion of flagged patients marked as "No" grows 50%, consider reconfiguring the pipeline (e.g. changing the prompt) or retirement. | Data Science team reviews Databricks Dashboard monthly and Analytics performs subgroup analysis yearly. | 1. Total number of flagged patients (weekly)<br>2. Of flagged patients, the number of patients co-managed by the SCM team (assignment of patients to the SCM treatment team in the EHR)<br>3. Count of SCM hospitalist admissions to per shift<br>4. Generation cost over time<br>5. Qualitative staff feedback | Monitor trends manually with clinical team and business owners.<br><br>If the proportion of flagged patients seen by SCM (metric 2) falls below 50% or qualitative tool usefulness falls below desired levels (metric 5) given cost (metric 4), consider reconfiguring the AI system or retirement. | Business owner reviews report 3 months post-deployment, and annually thereafter.<br><br>Analytics creates specific ad hoc subgroup analysis yearly |
| **AI-augmented Notes Review for Surgical Site Infections (SSIs)**<br><br><u>System details:</u> An LLM-powered workflow for identifying patients who may have experienced a surgical-site infection after surgery to assist in manual chart review for mandatory reporting.<br><br><u>Workflow details:</u> Used by quality specialists to streamline the review of patients with suspected associated infections. | 1. Number of eligible patients per week/month<br>2. Number and proportion of patients with suspected SSIs (positive scores)<br>3. Number of inference errors/failures | <u>Real-time support:</u> For metric 3, fire an alert if there are any inference errors for any run. Respond to alerts and resolve pipeline or inference issues<br><br><u>Governance review:</u> For metric 1, send an alert if the number of eligible patients increases outside a 75-125% band around historical values. For metric 2, review with clinical stakeholders if the number of flagged patients regularly exceeds review capacity. | <u>Real-time support:</u> Integration team responds to alerts in real-time.<br><br><u>Governance review:</u> Data Science team reviews Databricks Dashboard monthly. | Calculate the number and proportion of flagged (and a sample of unflagged) patients in each human feedback category:<br>• Yes, SSI suspected<br>• No, SSI not suspected<br><br>Annually, generate a report breaking down the above into patient subgroups identified during ethics assessment. | Investigate when...<br>• Precision (proportion of flagged patients with an SSI confirmed by a human reviewer) falls below the acceptable minimum (10%).<br>• Sensitivity falls below the acceptable minimum (95%).<br><br>If this occurs, consider reconfiguring the pipeline (e.g. changing the prompt) or retirement. | Data Science team reviews Databricks Dashboard monthly and Analytics performs subgroup analysis yearly. | 1. Number/proportion of non-flagged patients (those only requiring optional manual review) per month<br>2. Generation cost<br>6. Qualitative staff feedback on tool usefulness | If the proportion of patients not requiring manual review (metric 1) falls below 30% or qualitative tool usefulness falls below desired levels (metric 3) given cost (metric 2), consider reconfiguring the AI system or retirement. | Monitor trends manually with hospital acquired infection (HAI) quality team and Inpatient Applications team.<br><br>Impact metrics are reviewed 3 months post-implementation, and yearly after that. |
| **ChatEHR Interactive User Interface (UI)**<br><br><u>System details:</u> An EHR-embedded, secure chat | 1. Daily/weekly number of chat sessions/queries<br>2. Number of daily active users | <u>Real-time support:</u> Trigger alerts for sustained spikes in errors/timeouts or failures in data | <u>Real-time support:</u> Engineering/ integrations team responds | 1. Task mix over time derived from log-based task classification | Manual, *ad hoc* review of metrics 1-3 by the ChatEHR product team, | Data science and ChatEHR product team review performance | 1. Overall usage: Number of sessions, patient records, and queries (by | Reassess utility when adoption plateaus or declines; or | Data science and ChatEHR product team review impact |
| interface that enables clinicians and staff to “chat” with a patient’s longitudinal health record assembled in real time.<br><br>Workflow details: Used within the EHR for information retrieval, longitudinal summarization, and information synthesis after the completion of a short training module emphasizing the verification of outputs and recommended prompting practices. | 3. Number of unique patient records queried<br>4. Response time and latency (for data retrieval and LLM response generation)<br>5. Error and timeout rates<br>6. Token counts and LLM routing telemetry<br>Data sources used (notes, labs, medications, diagnostic reports, etc.) | retrieval. Route to engineering/integrations on-call team for immediate triage.<br><br>Governance review: Manual review of system integrity dashboards by ChatEHR product team, business owner, and technical leads to identify persistent latency degradation or failures warranting corrective action. | to alerts in real-time.<br><br>Governance review: Data science team reviews metric dashboard monthly. | 2. User feedback signals: rate of positive/ negative feedback and thematic categorization of free-text error reports<br>3. Under development: scalable detection of unsupported claims in generated responses | business owner, and technical leads.<br><br>Data science team, in collaboration with researchers at the School of Medicine, develop benchmarks and evaluation methodology for high-frequency tasks. | dashboards monthly. | department and user role)<br>2. User adoption: number of users active in 2 or more consecutive weeks<br>3. Cost: total tokens processed and LLM API spend, including average cost per query/session | when costs rise disproportionately to use | dashboards quarterly. |
| **(LLM-generated) Draft Denial Appeal Letters (Hospital Billing)**<br><br>System details: An LLM-powered workflow for drafting an explanation for why a payer should reverse its decision to deny payment for services based on clinical documentation.<br><br>Workflow details: This AI system will be used to decrease time and effort required to assemble the clinical basis for denial appeal in a denial appeal packet. | 1. Number of generated drafts<br>2. Number of generation errors/warnings | Real-time support: Applications responds to alerts to remediate pipeline or inference issues<br><br>Governance review: TDS Applications manually reviews metrics 1-2 using built-in Epic Model Feature Management Dashboard on an *ad hoc* basis and reports to business owner | Real-time support: Applications team responds to alerts in real-time.<br><br>Governance review: Application team reviews metric dashboard monthly. | 1. In-workflow user feedback (thumbs up/down based on perceived quality of the draft appeal letter)<br>2. Acceptance and rejection rates of the draft appeal letter (based on “Copy”, “Copy without References”, or unused draft) | Alert when...<br>• Negative user feedback rate exceeds 50% of generated drafts<br>• Utilization (“Copy” or “Copy without References”) decreases below 50%.<br><br>If either of these occur, consider reconfiguration or retirement. | TDS Applications team monitors user feedback and utilization dashboards monthly. | Revenue cycle KPIs for impact monitoring include the following:<br><br>1. Cost to collect ratio<br>2. Cost to generate draft denial appeal<br>3. Number of denials worked<br>4. Denial recovery rate,<br>5. Total posted amount from insurance payments<br>6. Insurance collection ratio<br>3. Number of (weekly/monthly) active users | Review key metrics before and after AI system implementation on an *ad hoc* basis with business owner | Applications team will review impact metrics at 3 months post-implementation and yearly afterwards. |
| **(LLM-generated) Draft Denial Appeal Letters (Professional Billing)**<br><br>System details: An LLM-powered workflow for drafting an explanation for why a payer should reverse its decision to deny payment for services based on clinical documentation.<br><br>Workflow details: This AI system will be used to decrease time and effort required to assemble the clinical basis for denial appeal in a denial appeal packet. | 1. Number of generated drafts<br>2. Number of generation errors/warnings | Real-time support: Applications responds to alerts to remediate pipeline or inference issues<br><br>Governance review: TDS Applications manually reviews metrics 1-2 using built-in Epic Model Feature Management Dashboard on an *ad hoc* basis and reports to business owner | Real-time support: Applications team responds to alerts in real-time.<br><br>Governance review: Application team reviews metric dashboard monthly. | 1. In-workflow user feedback (thumbs up/down based on perceived quality of the draft appeal letter)<br>2. Acceptance and rejection rates of the draft appeal letter (based on “Copy”, “Copy without References”, or unused draft) | Alert when...<br>• Negative user feedback rate exceeds 50% of generated drafts (with a minimum of 5% generated drafts with feedback)<br>• Utilization (“Copy” or “Copy without References”) decreases below 50%.<br><br>If either of these occur, consider reconfiguration or retirement. | TDS Applications team monitors user feedback and utilization dashboards monthly. | Revenue cycle KPIs for impact monitoring include the following:<br><br>1. Cost to collect ratio<br>2. Cost to generate draft denial appeal<br>3. Number of denials worked<br>4. Denial recovery rate,<br>5. Total posted amount from insurance payments<br>6. Insurance collection ratio<br>Number of (weekly/monthly) active users | Review key metrics before and after AI system implementation on an *ad hoc* basis with business owner | Applications team will review impact metrics at 3 months post-implementation and yearly afterwards. |

# Appendix 1

## Monitoring Plan Template

| **Ai System Monitoring Recommendation Summary: [PROJECT NAME] - Stanford Health Care** |
|---|
| **Use case description**<br><br>The AI [tool/system] being assessed, [name], is designed to [describe intended purpose]. The [clinical/operational] workflow it is to be integrated into aims to [briefly describe workflow]. |
| **Recommendation**<br><br>In collaboration with business owners, [list business owners here], and SHC employees who will take action based on [name] output, we recommend developing a **monitoring plan**.<br><br>Monitoring is the measurement of certain properties and the criteria to respond of deployed AI tools. Monitoring involves evaluating the observed impact on an AI-augmented workflow during and after deployment including regular assessment of both technical and operational aspects. A key part of monitoring is a plan of action including the defined metrics, frequency of review, and responsible individuals. Criteria for decisions need to be outlined ranging from debugging the pipelines and systems hosting the model if an output is not produced, to retraining the model if performance dips below an allowable threshold, to workflow interventions if user adherence is too low. There are three aspects of a deployed AI tool that will need to be monitored:<br><br>• **System integrity monitoring** ensures that the model functions correctly and produces an output (i.e., it "runs"). Key considerations include inference-time errors or warnings, connectivity, and the integrity of data pipelines to and from the model. Metrics in this category measure uptime, latency, errors, and outages.<br>• **Performance monitoring** assesses whether the model is correct by evaluating accuracy, positive predictive value (PPV), drift, and other performance-related metrics. Surrogate or proxy outcomes may also be used to gauge effectiveness.<br>• **Impact monitoring** focuses on whether the model's insights lead to the desired actions and outcomes. This includes tracking workflow adoption and adherence, gathering user feedback, and measuring impact. Operational metrics assess user adoption, value realization, and overall implementation success.<br><br>We recommend identifying *thresholds for system integrity and performance outputs* that should trigger retraining or retirement of [name]. We also recommend identifying a *minimum frequency of [intended impact-related event(s)] to support continued use*, and a *maximum frequency of [unintended impact-related or safety events] to support retirement*. Lastly, we recommend developing processes to guide the relevant [clinical/operational] workflow in absence of [name], should retirement be necessary.<br><br>The table below indicates the necessary level of detail when developing a monitoring plan. |
| |
| **System Integrity – related to AI tool infrastructure (uptime, latency, errors, outages etc.)** |

| **Metric** | **Tool/Alert Mechanism** | **Cadence** | **Responsible Party** | **Plan of Action** |
|---|---|---|---|---|
| For example: Input and Output<br><br>Errors, Warnings, Records scored per model version, and Feature category prevalence, missingness, median value | For example: Epic Model Feature Management;<br><br>Alert triggers to email | For example: Monthly | Identify specific team and team member name(s) | For example: TDS application analyst monitors notification events via email and categorize the errors. |
| **Performance – related to AI tool accuracy and quality (accuracy, PPV, sensitivity etc.)** | | | | |
| **Metric** | **Tool** | **Cadence** | **Responsible Party** | **Plan of Action** |
| For example: Specificity, Sensitivity, AUROC, AUPRC, PPV, C-Statistic, Model Flag Rate | For example: Radar Dashboard;<br><br>Specific ad hoc evaluation | For example: Monthly | Identify specific team and team member name(s) | For example: Investigate when model's performance metrics deviate from 75 – 125% of model validation performance. If model's performance has deviated 3 times in a year or more, retrain or retire. |
| **Impact – related to AI tool user adoption, value realization** | | | | |
| **Metric** | **Tool** | **Cadence** | **Responsible Party** | **Plan of Action** |
| For example: Readmission rate | For example: Readmission MGT Tableau Dashboard | For example: 3 months post-implementation, and yearly after that. | Identify specific team and team member name(s) | For example: Informaticist will monitor and report updates to business owner. Users may submit Helpdesk incidents in SNOW. |
| **References**<br><br>[ADD] | | | | |
| **Authors and Contributors**<br><br>[ADD] | | | | |